\documentstyle[11pt,psfig]{article}

\begin{document}

\titlepage

\def\a{\alpha}
\def\b{\beta}
\def\g{\gamma}
\def\d{\delta}
\def\e{\epsilon}
\def\x{\times}
\def\om{\omega}
\def\k{\kappa}
\def\D{ {\cal D}}
\def\rg{r_g}
\def\beq{\begin{equation}}
\def\eeq{\end{equation}}
\def\beqa{\begin{eqnarray}}
\def\eeqa{\end{eqnarray}}
\def\mxth{\mathsurround=0pt }
\def\xversim#1#2{\lower2.pt\vbox{\baselineskip0pt \lineskip-.5pt
\def\ialign{$\mxth#1\hfil##\hfil$\crcr#2\crcr\sim\crcr}}}
\def\simgr{\mathrel{\mathpalette\xversim >}}
\def\simle{\mathrel{\mathpalette\xversim <}}
\def\O{{\cal O}}
\def\L{{\cal L}}

\def\ra{\rightarrow}
\def\lag{Lagrangian}
\renewcommand{\thesection}{  }
\renewcommand{\theequation}{\arabic{equation}}

\begin{flushright}
TUM-HEP-209/94\\
UPR-644-T/94\\
MPI-PTh-95/34
\end{flushright}
\vspace{1ex}

\begin{center} \bf

SPIN--$\frac{1}{2}$ PARTICLE IN  GRAVITATIONAL FIELD OF
A ROTATING BODY
      \rm
\vspace{5ex}

{\bf Zygmunt Lalak$^{\dagger,\S}$, Stefan Pokorski$^\ddagger$,
Julius Wess$^\ddagger$}

\vspace{0.5cm}
 $^\dagger$ Physik Department \\
       {\em Technische Universit\"at M\"unchen} \\
       {\em D--85748 Garching, Germany}

\vspace{0.5cm}
 $^\ddagger$ Max Planck Institut f\"ur Physik \\
       {\em Heisenberg Institute}\\
       {\em D--80805 M\"unchen, Germany}

\vspace{0.5cm}
 $^{\S}$ Department of Physics \\
       {\em University of Pennsylvania} \\
       {\em Philadelphia, PA 19104}

\vspace{3ex}

ABSTRACT
\end{center}
Effective Lagrangian describing gravitational source spin-particle spin
interactions is given. Cosmological and astrophysical consequences of
such interaction are examined. Although stronger than expected,
the spin-spin interactions do not change any cosmological effect observed
so far. They are important for background primordial neutrinos.

\vspace{3ex}
\begin{flushleft}
TUM-HEP-209/94\\
UPR-644-T/94\\
MPI-PTh-95/34\\
December 1994
\end{flushleft}

\newpage
\noindent{\large \bf Introduction~} 
It is not uncommon to find in the Universe
rotating massive objects. This rotation may be rather slow, like
in the case of Earth,
or relatively rapid, like that of some neutron stars.
Spinning  of the source changes the resulting gravitational field
and introduces new with respect to the case of simple static sources,
angular momentum dependent, gravitational forces.
We find it interesting to study and clarify the status of
the source-spin dependent gravitational
interactions between source and particles travelling through its
field. We consider a simplest case of
nonzero-spin particle, a spin-$1/2$ fermion. To be specific,
in this note we shall discuss in some detail  neutrino interacting
with the spinning Sun, but in fact our effective Lagrangian introduced in
Section 2. is a general one, valid for any kind of spin-half fermion
and any rotating source, like a pulsar or a rotating black hole.

\noindent{\large \bf 1.Gravitational field of a rotating body~} 
The gravitational field of a spinning sphere
 of mass M and
angular momentum $\vec{J} = M \vec{a}$ is described by the Kerr
metric, which is an exact solution to the Einstein equations.
Since we are going to apply methods of Minkowski space field
theory to interactions of spin $\frac{1}{2}$ fermions with the
spinning background, it is meaningful and sufficient to consider
the asymptotic form of the Kerr metric obtained in the limit
$\frac{1}{r} \rightarrow 0$. As we are going to consider the
effects of rotation, we give the asymptotic form of the Kerr
metric up to terms $(\frac{\rg}{r})^2$ and $(\frac{\rg}{r} \,
\frac{a}{r})$, where $\rg=2M/M^{2}_{Planck}$ is the
Schwarzchild radius,  (we use units and notation
of \cite{land})
\beqa
g_{00} &=& 1 - \frac{\rg}{r} +(\frac{\rg}{2 r})^2 \nonumber\\
g_{ij} &=& -\delta_{ij} (1+\frac{\rg}{r} +\frac{1}{2}
(\frac{\rg}{r})^2)\nonumber\\
g_{0j} &=& \frac{\rg}{r^3} \om_j\nonumber \\
\om_j &=& (\vec{a} \times \vec{x})_j
\label{kerr}
\eeqa
The reference frame has been fixed to the axis of rotation of
the body and the metric is written in the so-called isotropic
coordinates (the coordinate system in which the asymptotic
Schwarzchild metric assumes the diagonal form).
One should note that the
exact Kerr metric gives the upper limit on the radius $a$, $a <
\rg /2$ (cf. \cite{land}).

The asymtotic metric (\ref{kerr}) can also be obtained,
 without any
reference to the Kerr metric, just by solving the linearized
(in weak field approximation) Einstein equations \cite{land}.
Our considerations are valid for any metric which has the
asymptotic form (\ref{kerr}).
The vierbeins and Christoffel symbols corresponding to the
metric (\ref{kerr}) are listed in the Appendix.

\noindent{\large \bf 2.Effective lagrangian for spin--$\frac{1}{2}$
particle in
the spinning background~}  
The general invariant coupling of spin 1/2
fermions to gravity is given by (conventions we use are those of
Bjorken and Drell \cite{bjo})
\beq
\L = \sqrt{-g} (i \bar{\psi} \gamma^a D_a \psi - m \bar{\psi}
\psi )
\label{lag0}
\eeq
where $D_a = e^{\mu}_{a} (\partial_{\mu} + \frac{1}{2}
\Sigma^{cb} e^{\nu}_{c} e_{b \nu ; \mu}), \Sigma^{cb} =
\frac{1}{4} [\gamma^c, \gamma^b], \, e_{b\nu;\mu} =
(\delta^{\gamma}_{\nu} \partial_{\mu} -
\Gamma^{\gamma}_{\nu \mu}) e_{b \gamma}$. The $a,b,c,...$ are
flat indices and $\a, \b, \g,...$ are curved space indices.

In consistency with the metric (\ref{kerr}), valid, to repeat,
in weak field and slow rotation limit, we retain in the
lagrangian (\ref{lag0}) only terms of up to the order
$O((\frac{\rg}{r})^2)$ and $O(\frac{\rg}{r} \frac{a}{r})$.
Then the final form of the relevant Dirac operator is
\beqa
 D &=& \g^a D_a = (i \g^a \partial_a -m ) + i \g^0
\frac{\rg}{2 r} \partial_0 + \nonumber \\
& +& \frac{i f}{2} \g^i \partial_i - \frac{i}{4}
\frac{\rg^2}{r^4} \vec{x} \vec{\g} + \nonumber \\
 &-& \frac{\rg}{r^3} \vec{\g} \vec{\om} \partial_0 +\nonumber \\
&+ & \frac{\rg}{4 r^3} \g^5 \vec{\g} \vec{a} - \frac{3 \rg}{4
r^5} \g^5 (\vec{\g} \vec{x}) ( \vec{a} \vec{x} )
\label{dirac}
\eeqa
where $f=-\rg / r -1/2 (\rg /r)^2 $. The operator (\ref{dirac})
is not explicitly Lorentz-invariant (but it is $O(3)$
invariant). In fact, to write explicitly the interactions with
the background we had to choose the specific coordinate system,
the one where the source of the background stays at the origin.
Another issue is the gauge invariance (the invariance under
small coordinate reparametrizations, e.g. the change from
isotropic to Schwarzchild coordinates). One can check that as
long as one considers only terms which are lowest order in
expansion parameters (i.e $O(\frac{\rg}{r}), \, O(\frac{\rg}{r}
\frac{a}{r}) $), and one restricts oneself to small gauge
transformations, the changes in the interaction terms are higher
order ones. \\
Various terms in (\ref{dirac}) have a straightforward interpretation.
The first term is the usual ``flat'' Dirac operator, the second term
describes the central attractive force and the spin-orbit interaction.
The terms in the second line don't cause any spin-flip as they are
spin independent operators. The term from the third line describes
the interaction of the orbital
angular momentum of the particle with the spin of the
background\footnote[1]{One can  see that
the leading order in the nonrelativistic expansion of
the hamiltonian density corresponding to this
term is proportional to $\vec{a} \vec{L} E \rg /((E+m) r^3)$.}, and
the last line contains the tensorial operator describing
spin-dependent gravitational interactions of the particle in question.

\noindent{\large \bf 3.Effects of interaction with gravitational field~} 
In this section we estimate the magnitude of
several physical effects on spin 1/2 particle interacting with
gravitational background. We are primarily interested in the
effects of the spin-spin interactions, hence we neglect the
terms involving the angular momentum operator. Physically,
we can imagine the particle  travelling (or
emitted) radially, at some angle $\theta_i$ with respect to the
rotation axis. One should notice at this point that the
contribution to the relevant cross sections coming from orbital
momentum interactions can be added incoherently to the
spin--spin cross sections at the level of tree-graph processes.

Secondly, we are of course interested in the coherent
interaction of the particle spin with the total, macroscopic
spin of the body, as this can in principle enhance the
gravitational strength interactions which we consider. Hence,
we have to assume some reasonable ultraviolet momentum cut-off
for the allowed range of momentum transfer during the scattering
event. In the following we assume as the cut-off the inverse
Schwarzchild radius
\beq
q^2 < \frac{1}{\rg^2}
\label{cut}
\eeq
where the momentum transfer is $q^2 = 4 p^2 \cos^2 (\theta /2)$,
$p$ being the momentum of the incoming particle and $\theta$ the
scattering angle.
We note that in any case our metric (\ref{kerr}) is not
expected to hold below the Schwarzchild radius.

We shall discuss the following effects:
\begin{description}
\item[1)] Energy level shift for a particle with positive and negative
helicities;
\item[2)] The cross section for spin flip in the process of scattering
in the spinning background;
\item[3)] The alignment of spin along the direction of the rotation
axis.
\end{description}
To estimate the effects (2)--(3) we imagine the particle to be
scattered from an asymptotic {\em in} state to an asymptotic
{\em out} state (both corresponding to the flat Minkowski background far
from the source) and apply the standard rules for calculating the
scattering cross sections in the external field described by
\beq
\delta \L =  \frac{\rg}{4 r^3} \g^5 \vec{\g} \vec{a} -
 \frac{3 \rg}{4
r^5} \g^5 (\vec{\g} \vec{x}) ( \vec{a} \vec{x} )
\label{spin}
\eeq
Although this may be not more than a crude approximation to the
effects expected for a particle e.g. produced in the rotating
body, hopefully it provides the correct order of magnitude
estimate for the actual interaction.

Let us note that interaction (\ref{spin}) resembles closely the interaction
between a magnetic dipole moment and the magnetic field produced by another
dipole $\vec{m}$
\beq
\delta H_{mag} = -\vec{\mu} \; \vec{B}_m
\label{mag}
\eeq
where $\vec{B}_m (\vec{x})= \frac{3 \vec{n} (\vec{n} \vec{m}) - \vec{m}
}{|\vec{x}|^3}$. In the present case we can write (\ref{spin}) as
\beq
\delta H = - \vec{S} \vec{\cal B} = - \frac{1}{2} \vec{\Sigma} \vec{\cal B}
\label{anal}
\eeq
where $\vec{\cal B} (\vec{x}) =  \frac{3 \vec{n} (\vec{n} \vec{J}) - \vec{J}
}{|\vec{x}|^3}$ with angular momentum of the background $\vec{J} = M \;
\vec{a}$ corresponding to the magnetic moment $\vec{m}$ from (\ref{mag}).
For a massive Majorana neutrino, which is not allowed to have static
magnetic or electric dipole moments, the interaction
(\ref{spin}),(\ref{anal}) is the only possible dipole-dipole type interaction
(up to the order of magnitude considered here).\\
\noindent{\em Shift of energy levels between opposite chirality
states~}\\
We calculate the shift of energy levels
\beq
\delta E = < \bar{\Psi} | -\gamma^0 \delta \L | \Psi>
\eeq
with $\delta \L$ given by (\ref{spin}) and the wavefunction of
the fermion given by a well localized packet
\beq
\Psi(p,\lambda) = \int_{x} f(x,p) u(p,\lambda)
\eeq
We assume that the function f is sufficiently well localized to
pull the factors of $1/r$ outside the space integral. Spinors
$u$ are normalized as $\bar{u} u =2 m$
hence the normalization of the packet profile f is $ \int d^3 x
\bar{f} f = \frac{1}{2 E} $.
Let's assume that the direction of $\vec{a}$ coincides with the
z-axis. Then the energy shifts $\delta E^{+,-}$ of the states
with positive (negative) helicity are
\beq
\delta E^{(+)} = - \delta E^{(-)} = \frac{\rg a}{4 r^3}
(\cos(\theta) - 3 \frac{\vec{a} \vec{x}}{a r} (\frac{x^1 }{r}
\sin(\theta)  + \frac{x^3 }{r} \cos(\theta) ))
\label{shift}
\eeq
In the specific case of ``radial'' emission from the rotating
star, $\frac{\vec{x}}{r} = \frac{\vec{p}_i}{r} = \vec{n}$,
\beq
\delta E^{(+)} = - \delta E^{(-)} = - \frac{\rg a}{2 r^3}
\cos(\theta)
\label{shrad}
\eeq
This can be interpreted as the mass shift of the helicity
eigenstates, $\delta m = \frac{\rg a}{ r^3} |\cos(\theta)|$.
In the case of the Sun, which will be our standard example,
this mass shift is
\beq
\delta m = 0.7 \times 10^{-27}
(\frac{r}{R_{\odot}})^{-3} |\cos(\theta)| \, eV
\eeq
For $r=r_{g \odot}$ one gets $\delta m = 8.9 \times 10^{-12}
|\cos(\theta)| \, eV $. \\
\noindent{\em The scattering cross section~}\\
The differential cross section for scattering in
the field (\ref{spin}) is given by
\beq
d \sigma = \frac{1}{16 \pi^2} |M_{fi}|^2 d \Omega
\eeq
where
\beq
M_{fi} = < f|\delta \L| i>
\eeq
The fourier transform of (\ref{spin}) is
\beq
{\cal F} (\delta \L) = \int_{\rg}^{\infty} d^3 x e^{i \vec{q} \vec{x}}
\delta \L = T(q) \frac{\rg a}{8} \g^5 \vec{\g} \vec{\beta}
\label{fourier}
\eeq
with $\vec{\beta} = 3 (\vec{n}_a \vec{n}_q) \vec{n}_q - \vec{n}_a$
and $T= 4 \pi (\frac{\cos(q \rg )}{(q \rg)^2} - \frac{\sin(q \rg)}{(q
\rg)^3})$, $\vec{n}_a = \vec{a}/a$, $\vec{n}_q =\vec{q}/q$.
T approaches the constant value $-4 \pi /3$ for small $q$ and
oscillates approaching zero
 as the function of the
natural variable $q \rg = 2 p \rg \sin(\theta /2 )$. As we do not
want to penetrate    the inside of the Schwarzchild radius, we restrict the
allowed range of $q$ by the condition $q \rg <1$.

We are now ready to calculate the differential cross section for transition
between different spin states. Let $\theta_{i(f)}$ denote the angle between
the incoming (outgoing) momentum and the axis of rotation, and $\theta$ the
scattering angle.
The final formulae for helicity flip ($\sigma_{LR}$), no-flip ($\sigma_{LL}$)
and the total spin-spin interaction ($\sigma_{TOT}$) cross-sections are
\beqa
\frac{d \sigma_{LR} }{d \Omega}& =& \frac{1}{2^{8} \pi^2} 2 a^2 m^2 \rg^2
|T|^2 ( \frac{3}{2} (\cos(\theta_f) -\cos(\theta_i))^2 \nonumber \\
{}& {}& (1+2 \cos^2 (\theta /2)) + 2 \cos^2 (\theta /2) - 2 \cos (\theta_f)
\cos (\theta_i ))  \label{lrf} \\
\frac{d \sigma_{LL} }{d \Omega}& =& \frac{1}{2^{8} \pi^2} 2 a^2 m^2 \rg^2
|T|^2 \frac{E^2}{m^2} ( 1 - \cos(\theta) +  2 \cos (\theta_f)
\cos (\theta_i ))  \label{llf} \\
\frac{d \sigma_{TOT}}{d \Omega}&=& \frac{d \sigma_{LR}}{d \Omega} +
\frac{d \sigma_{LL}}{d \Omega}
\label{crosss}
\eeqa
where $\Omega$ is the solid angle around the direction of the scattered
particle and the angles $\theta_i , \; \theta_f , \; \theta$ and the azimuthal
angle $\phi$ are related through $\cos (\theta_f )= - \sin (\theta_i )
\sin ( \theta ) \cos (\phi ) + \cos ( \theta ) \cos (\theta_i )$.

For the benefit of the further discussion we have  collected in the Appendix
the values of the parameters entering the cross section formulae relevant for
the case of the Sun.

\noindent{\large \bf 4.Discussion and Conclusions~} 
Let us discuss the spin-flip effect first.
The differential cross section $d \sigma_{LR}$ we have got has
the general form
\beq
d \sigma_{LR}= \k T^2(\rg q) A(\theta_i , \theta, \theta_f) d
\Omega
\label{lr}
\eeq
where $\k = 8 \times 10^{-4} a^2 \rg^2 m^2$, $q \rg = 2 p \rg
\sin(\theta /2)$ and $A$ is some function of angular variables only.
Moreover, when calculating the total cross
sections we
have to restrict our angular integration to scattering angles
smaller than $\theta_{max}= \frac{1}{p \rg}$.
The integrated spin-flip scattering cross section as
a function of the variable $R=2 p \rg$ for fixed values of the
angle $\theta_i$ can  be found numerically.
Using the analytic formula for T(q) and formula  (\ref{lrf}) one
can   find an approximate expession for the magnitude of the integrated
spin flip cross
section
\beq
\sigma_{LR} = \left \{ \begin{array}{cc}
10^{-13} (p/ GeV)^{-2} (a GeV)^2 (m/keV)^2 GeV^{-2}& p \rg > 1 \\
10^{-13} (\rg GeV)^2 (a GeV)^2 (m/keV)^2 GeV^{-2}& p \rg < 1
\end{array} \right.
\label{sim}
\eeq
(this is assuming $a \approx \rg$). Numerical values of the
gravitational spin flip cross section should be compared to the average
weak spin flip cross section \cite{gem} assuming that the spin $1/2$
particle in question interacts weakly (like a massive neutrino)
\beq
\sigma_{Z, flip} = 1.6 \times 10^{-23} (m/keV)^2 \; GeV^{-2}
\label{weak}
\eeq
Let us check when the gravitational spin flip becomes comparable
to the weak spin flip
\beq
\sigma_{LR} / N =\sigma_{LR} /(M/m_{proton}) \geq \sigma_{Z flip}
\label{5.4}
\eeq
($N$ is the number of scattering centers in the body.)
Using the formula (\ref{sim}) one can   obtain the
condition
\beq
M \geq {\cal M} = 10^{47} \; GeV
\label{5.5}
\eeq
(with $p < \frac{1}{\rg}$),
which  is fullfilled in the case of the Sun (however, the neutrinos actually
coming from the Sun have much larger momenta), and the other condition
\beq
M \geq 10^{-10} (M_{Planck}/ GeV)^{4} (p/GeV)^2 \, GeV
\label{5.6}
\eeq
which is relevant for relativistic fermions. This last condition
is also   fulfilled in the case of the Sun as long as,
approximately,
$p \leq 10\, keV$ (which is still to small to be seen).
One should also note that according to (\ref{sim}) for larger
$p$ the cross section falls off like $m^2/ p^2$.
A possibility for the larger cross section is hidden in the
factor $a^2$ ($\sigma_{LR} \approx \frac{a^2 \, m^2}{p^2}$),
but the relation $a > \rg $ implies presence of the naked singularity
for a rotating blackhole, and also leads  beyond the range of
validity of the perturbative expansion of the metric employed
in our calculation. From the point of view of the analogy (\ref{anal})
the conditions on the mass $M$ of the source or/and on its angular
momentum ($M \; \vec{a}$) correspond to determination of critical
``magnetic fields'' which make the effect of gravitational dipole-dipole
interaction important in comparison with other forces in given
contexts.

To discuss the possibility of the alignment of spin of the
scattered particles along the axis of rotation, let us define
the average angle between the spin of the scattered particles
and the rotation axis (we work in helicity basis and assume that
the incoming beam of particles has a definite helicity)
\beq
<\cos(\theta_s )> = \frac{1}{\sigma_{TOT}} \int d \Omega
\cos(\theta_f) (\frac{d \sigma_{LR}}{d \Omega} - \frac{d
\sigma_{LL}}{d \Omega} )
\label{sppas}
\eeq
This average angle should be compared with the angle between the
spin of the incoming particle and the rotation axis.
We have to point out that in the relativistic case the effect of
the spin flip onto the spin alignment is negligible.
The no-flip cross section is related to spin-flip one through
the approximate relation
\beq
\sigma_{LR} \approx (1+\frac{p^2}{m^2})^{-1} \sigma_{LL}
\label{app}
\eeq
and in the ultrarelativistic case $\sigma_{LL} >> \sigma_{LR}$.
Thus, in this extreme case any observable deflection of spin outwards or
towards the rotation axis is due to the nontrivial distribution
of final momentum with respect to this axis, determined solely by the
no-flip
cross section.
However, as p falls closer to  m, both components of the cross
section do contribute, as discussed later.

Similarly, to discuss the alignment of the final momentum one
defines
\beq
<\cos(\theta_p )> = \frac{1}{\sigma_{TOT}} \int d \Omega
\cos(\theta_f) (\frac{d \sigma_{LR}}{d \Omega} + \frac{d
\sigma_{LL}}{d \Omega} )
\label{spass}
\eeq
which, again, has to be compared with $\theta_i$.
Let us note that, taking  into account (\ref{sppas}), (\ref{app}),
(\ref{spass}), in the ultrarelativistic
limit $<\cos(\theta_p )> = - <\cos(\theta_s )>$
Fortunately, one can write down approximate expressions for $<\theta_s>$ and
 $<\theta_p>$
valid when $p \rg > 1$, which covers most cases of practical
interest\footnote[2]{In the case of the Sun this inequality means $p > 10^{-10}
\;eV$.}
\beqa
<\cos(\theta_s )> &=& \cos(\theta_i) \; (-1 +\frac{2  \sin^2 (\theta_i)}{\sin^2
(\theta_i) +
\frac{E^2}{m^2} \cos^2 (\theta_i)}- \frac{0.24}{p^2 \rg^2}\; \nonumber \\
{ }&\times& \frac{1}{(
 \sin^2 (\theta_i) + \frac{E^2}{m^2} \cos^2 (\theta_i))^2} (4 \sin^4 (\theta_i)
+ ( \frac{E^2}{m^2})^2 \sin^2 (2 \theta_i) \nonumber \\
{ }&+& 2 \frac{E^2}{m^2} (2+ 2 \sin^2 (\theta_i) -9 \sin^2 (\theta_i)
\cos^2 (\theta_i))
) \label{spas}\\
<\cos(\theta_p )> &=& \cos(\theta_i) \; (1 + \frac{0.24}{p^2 \rg^2}\;
\frac{( \frac{E^2}{m^2}-1) \sin^2 (\theta_i) }{
 \sin^2 (\theta_i) + \frac{E^2}{m^2} \cos^2 (\theta_i)})
\label{aprsect}
\eeqa
(where using the approximate expressions one should take care that the
corrections are smaller than the leading contributions over the whole range
of $\theta_i$, the relevant condition being $r< p^2 \rg^2 /0.24$).

The distributions of $\theta_s$ and $\theta_p$
as functions of the angle $\theta_i$ between initial momentum and the rotation
axis
are shown in the Figure 1  for $x=E^2 / m^2$ equal to $4$ and $100$.
Figure 1(a) shows the case  when
the momentum of scattered particle lies in the low energy range i.e. is
negligible with respect to the planck scale,  Figure 1(b) shows
the case  when momentum is comparable
to the  planck scale.
It is clearly visible that the  average final spin gets deflected towards the
direction of the source spin, $\vec{a}$,
 when $\theta_i < \pi /2$ and outwards
when $\theta_i > \pi /2$. These deviations from the incoming direction of spin
become  smaller the more relativistic the particle is (the larger $x$ becomes),
and have maxima,
one for $\theta_i < \pi /2$ and one for $\theta_i > \pi /2$, which are
approaching  $\pi /2$ as the ratio $x$
grows. As seen from (\ref{aprsect}) the deflection of the momentum is a
second-order effect in the expansion we use, proportional to $p^{-2} \rg^{-2}$,
and
for typical momenta encountered in terrestial or solar physics it is
practically zero. To be able to draw lines visible on the picture 1(b) we have
assumed an exotic value of the order parameter, $p \rg = 6$
(which implies planck scale p and also very large m for chosen values of x).
One can see that  the outgoing momentum gets deflected towards
the direction of the source spin
\par
\centerline{\hbox{
\psfig{figure=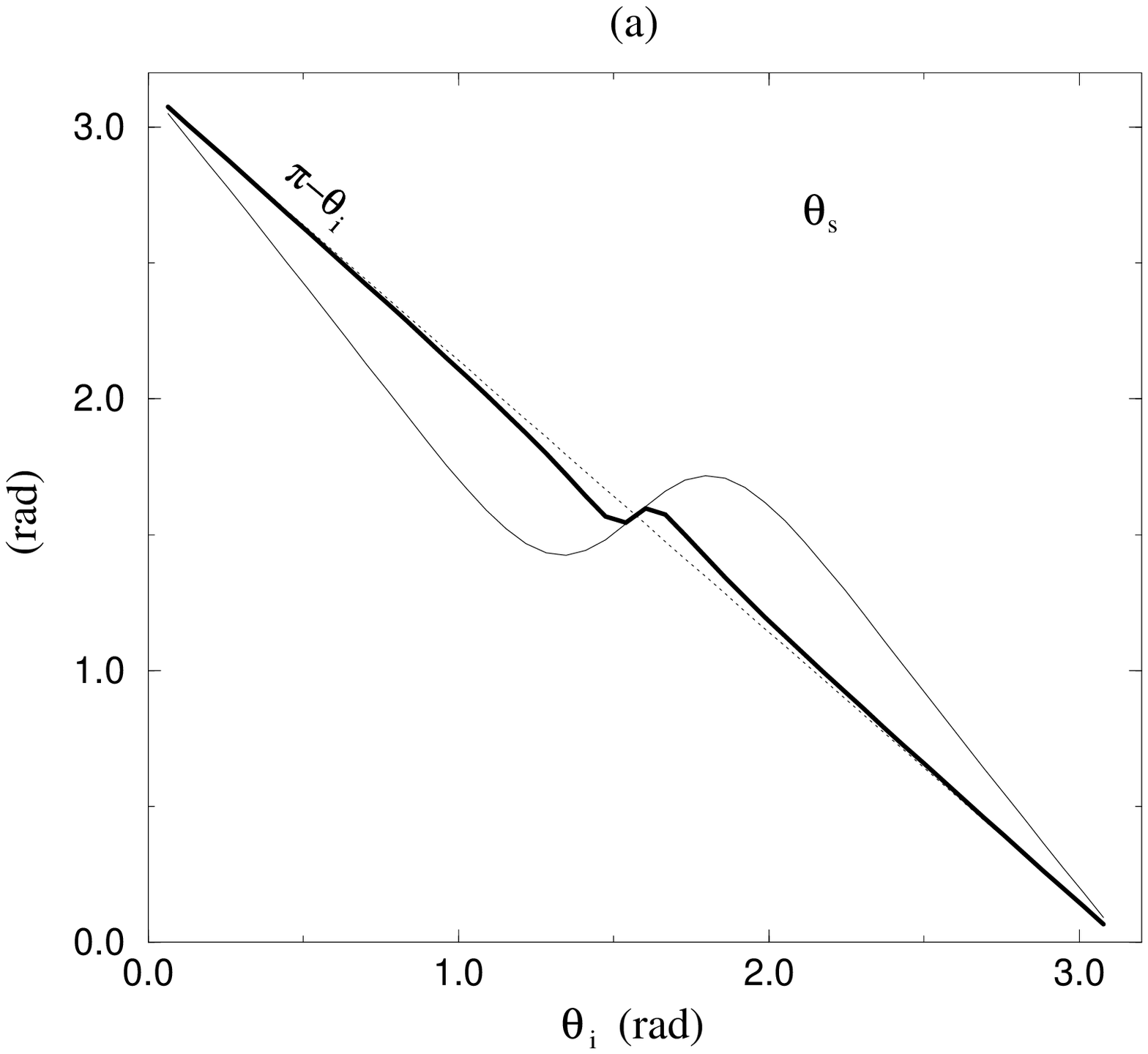,height=3.0in,width=3.0in}
\psfig{figure=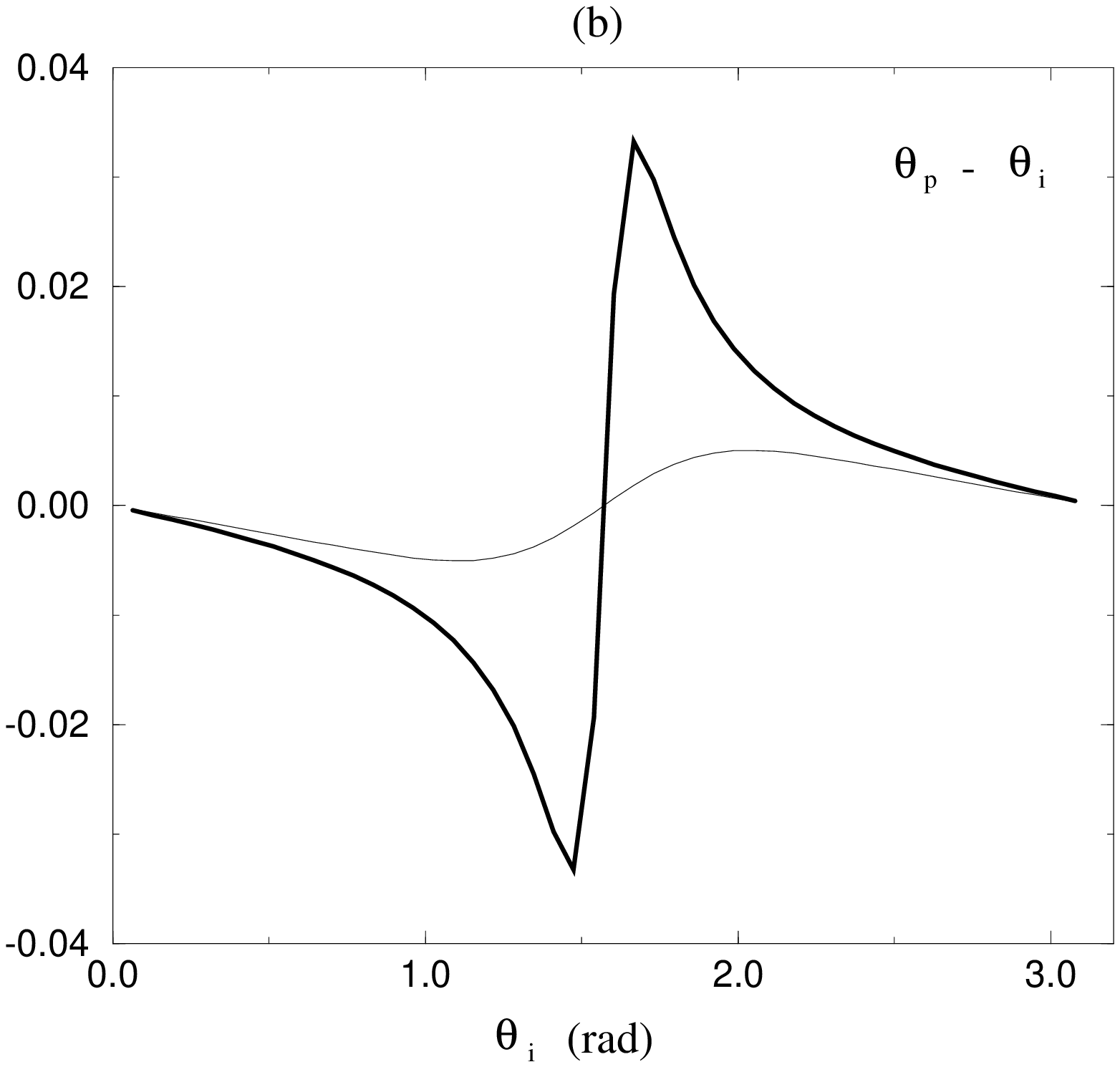,height=3.0in,width=3.0in}
}}
\par
{\small \em Figure 1. Distributions of the mean  angles  between
average outgoing spin, $\theta_s (\theta_i)$, average outgoing
momentum,  $\theta_p (\theta_i)$, and the rotation axis. Plots are given
for two values of the ratio $x=E^2 / m^2 $, $x=4$ (solid lines) and $x=100$
(thick solid lines).
Figure (a) shows the case of terrestial-scale momentum, when the $\theta_p$
deviation from $\theta_i$ is heavily suppressed and invisible on the picture.
Figure (b) shows the distribution of $\theta_p - \theta_i$ visible when
momentum enters the planck scale region -- it is obtained with  $p \rg =6$
 (initial spin polarization is assumed antiparallel to incoming momentum).}\\
\vspace{0.3cm}

\noindent{when $\theta_i < \pi /2$}
and away from the vector $\vec{a}$ when $\theta_i > \pi/2$.
Again, there are maxima in the deflection angle $| \theta_p - \theta_i |$ which
come closer to $\theta_i =\pi/2$ as $x$ grows. The visible tendency for
the deflection in momentum distribution in Figure 1(b) to grow with growing $x$
whereas
the deflection in spin distribution in Figure 1(a) gets suppressed for larger
$x$ stems
from the fact, that Figures 1(a) and 1(b) are drawn for vastly different values
of momentum p. Hence, the two distributions are determined by expressions which
are of different order in parameter $1/(p \rg)$ and  have different
x-dependence. The leading, $o(\frac{1}{(p \rg)^0})$ term in spin asymmetry
(\ref{spas}) behaves as $o(\frac{1}{x})$ when $x$ grows. But the nonleading,
$o(\frac{1}{(p \rg)^2}$ terms in both (\ref{spas}) and (\ref{aprsect})
contain $x$ both in numerators and denominators. They increase with $x$ and
in the limit of very
large $x$
they behave as $O(x^0 )$. In fact,  (\ref{spas}) and (\ref{aprsect})
become the same expression, up to a sign, in the limit $x \rightarrow \infty$
as they should in agreement with the remark after (\ref{spass}).

Finally, one should compare the magnitude of the effects due to the spin-spin
interaction with the effects of spin-orbit interaction, which arises due to
the existence of the second  term in the dirac operator (\ref{dirac})
as the standard Thomas-precession term. The total spin-flip cross section
caused by L-S interaction is   computed to be (we consider the example of
the Sun again\footnote[3]{For the Sun the spin-orbit interaction was considered
in ref. \cite{dass}})
\beq
\sigma_{LR-LS} = \frac{\pi E^2 m^2 \rg^2}{2 p^4} \log (2 p R_{\odot})
\eeq
The spin-orbit cross section falls down for large momenta
at the same rate as the spin-spin flip cross section
(like $\frac{m^2}{p^2}$) with similar numerical coefficients,
so for the relativistic neutrinos (fulfilling however the
conditions formulated above) the two effects can be comparable.
Also, the spin-orbit  cross section has an obvious peak at small momenta
whereas the spin-spin cross section approaches a constant value
as $p$ goes to $0$.
One has to admit that, in general, the spin-orbit interaction forms a
 significant background hiding the spin-spin effects. However,
there are certain kinematical conditions which are in favour of spin-spin
interaction, for instance the situation of quasi-radial emission of
neutrinos from the Sun, from cores of supernovae or neutron stars, when the
orbital angular momentum is naturally suppressed. \\

Let us discuss briefly possible cosmological implications of our results.
First, we have to say that as far as solar neutrinos are concerned,
that have  energies between $0.1-10 \;MeV$ and in typical models masses
between $10^{-3}$ and $1$ eV \cite{lang-sol}, our effect is subdominant
with respect to weak or magnetic moment spin-flipping interactions,
although it is not as dramatically small as one would be tempted to
claim naively. If there would be in the solar spectrum neutrinos with
energies smaller than approximately $10$ keV, then spin-solar-spin
interactions of these particles would be important.
The domain where spin-dependent interactions of neutrinos {\em are}
important, if there  are massive ones, is physics of background primordial
neutrinos, for review
cf. \cite{lang-pri}. These primordial neutrinos have today the average momentum
of $5.2 \times 10^{-4}$ eV and, with a source of the solar or larger mass,
they typically interact reasonably strongly via gravitational spin-spin
interactions as seen from (\ref{5.4}),(\ref{5.5}),(\ref{5.6}) -- at least
stronger than
 weakly. Of course, also in this case for the flipping
 to be comparable to no-flip interactions these particles should be
nonrelativistic, i.e. sufficiently massive.\\
Hence, one expects the primordial neutrino sea to be partially
polarized (spins aligned along the background angular momentum) in the vicinity
of the
Sun\footnote[4]{One should note at this point that, for instance, Sun moves
with respect to the
background with a velocity $v_S \approx 10^{-3}$. But, if neutrino masses are
in the typical model range,
i.e. at most of the order of a few eV, cf. \cite{lang-sol}, then in solar rest
frame they have momenta
at most of a few times $10^{-3}$eV, so they are still nonrelativistic if they
were so in the background frame.
The same applies to other kinds of nonrelativistic relics.}, and generally in
the
vicinity of any massive, rotating body in the Universe. The exact nature of the
final spin state of the
local neutrino sea would depend on the local kinematics, in particular on the
degree of anisotropy
of the momentum distribution of the neutrinos in the source's center of mass
rest frame.
However, at present we are not aware of any real or ``gedanken'' experiment
which can see and test the neutrino background. Similar conclusions hold
also for other fermionic primordial relics, in fact the effect should be
particularly important for massive warm or cold relics, like background
gravitinos, if they exist.\\
At last, let us discuss spin-spin interactions in the context of the supernova
physics \cite{gaemers-r}. As pointed out in the context of weak or magnetic
moment interactions if the spin flipping is to efficient, then the sterile
right-handed neutrinos stream freely from the supernova core and the supernova
cooling is too rapid. In fact in the case of the supernova SN 1987A the stream
of neutrinos was observed over a period of the order of $10$ seconds,
which gives a direct limit on the flipped neutrino luminosity from the
supernova,
$ L < 4 \times 10^{-22}$GeV${}^2$. We can   compute
the flipped neutrino luminosity due to the gravitational spin-flip
\beq
L_g \approx 1.8\; N_{\nu} \sigma_{LR}\; T^4
\eeq
where $N_{\nu}$ is the number of neutrino flavours considered.
If one demands the $L_g$ to be smaller than the limiting value quoted above and
taking the standard reference value for the temperature T, $T_o =30 \; MeV$,
then one obtains the limit
\beq
(\frac{a}{r_g})^2 (\frac{m}{keV})^2 (\frac{T}{T_o})^4 < 0.7 \times 10^{10}
\eeq
If one takes $a/r_g \approx 1$ then one gets an upper limit on
m, $m< 30 \; MeV$. This number coincides with similar limits coming from
weak and magnetic moment interactions. Unfortunately, our limit is in
fact weaker. The analysis of possible models for rapidly rotating pulsars,
\cite{friedman}, has shown that reasonable values of $a$ are rather $0.30-
0.34 \; r_g$, hence our limit probably
becomes an order of magnitude weaker.

In conclusion, we have described and examined the interaction of spin-one half
fermions with the spin of the local gravitational field due to rotation of
some massive body. These interactions, although more effective than one
would naively guess, do not seem to change any cosmological effect observed so
far.
They can become important in case of unusually fast rotating and massive
pulsars, and
they are important for background primordial neutrinos and for other
primordial fermionic relics from the Big-Bang epoch.

\vspace{1.0cm}
\noindent{\large \bf Acknowledgements~}
Authors thank Paul Langacker for interesting discussions about neutrinos,
and Georg Raffelt and Sidney Bludman for useful conversations about
neutron stars.\\
Work partially supported by Deutsche Forschungsgemeinschaft, Polish
Government's KBN and EEC grant ERBCIPDCT940034.
Z.L. was supported by A. von Humboldt Fellowship, and partially by
NATO grant CRG-940784.
\hspace*{.5cm}

\setcounter{section}{0}
\setcounter{equation}{0}

\renewcommand{\theequation}{A.\arabic{equation}}
\noindent{\large \bf Appendix~}
The vierbein which reconstructs the metric
(\ref{kerr}) through $g_{\mu \nu} = \eta_{\alpha \beta}
e^{\alpha}_{\mu} e^{\beta}_{\nu}$ we choose in the form
\beqa
e^{0}_{\mu}&=& (1-\frac{\rg}{2 r} - \frac{\rg^2}{8 r^2},
\frac{\rg}{r^3} \vec{\om}) \nonumber \\
e^1_{\mu}& =& (0, 1+\frac{\rg}{2 r} + \frac{\rg^2}{4 r^2},0,0)
\nonumber \\
e^2_{\mu}&=& (0,0, 1+\frac{\rg}{2 r} + \frac{\rg^2}{4 r^2},0)
\nonumber \\
e^3_{\mu}&=& (0,0,0, 1+\frac{\rg}{2 r} + \frac{\rg^2}{4 r^2})
\label{viel}
\eeqa%
The inverse vierbein is
\beqa
e^0_{\alpha}&=&(( 1-\frac{\rg}{2 r} - \frac{\rg^2}{4 r^2})^{-1},
-\frac{\rg}{r^3} \vec{\om}) \nonumber \\
e^1_{\alpha}&=&(0, ( 1+\frac{\rg}{2 r} + \frac{\rg^2}{4
r^2})^{-1},0,0) \nonumber \\
e^2_{\alpha}&=&(0,0, ( 1+\frac{\rg}{2 r} + \frac{\rg^2}{4
r^2})^{-1},0) \nonumber \\
e^3_{\alpha}&=& (0,0,0, ( 1+\frac{\rg}{2 r} + \frac{\rg^2}{4
r^2})^{-1})
\label{iviel}
\eeqa%
One can easily find Christoffel symbols for the metric
(\ref{kerr}), we list below the nonvanishing ones
\beqa
\Gamma^0_{0l}&=&\frac{1}{2 r^3} \rg x^l \nonumber \\
\Gamma^i_{00}&=&\frac{1}{2 r^3} \rg x^i \nonumber \\
\Gamma^0_{kl}&=&- \frac{3 \rg}{2 r^5} (x^l \om^k + x^k \om^l )
\nonumber \\
\Gamma^i_{0j}&=&\frac{\rg}{r^3} \e_{jki} a^k - \frac{3 \rg}{2
r^5} (x^i \om^j - x^j \om^i ) \nonumber \\
\Gamma^i_{kl}&=& \frac{1}{2} (f_{|i} \delta_{kl} - f_{|l}
\delta_{ik} - f_{|k} \delta_{il} )
\label{chris}
\eeqa%
where $f_{|i}= \frac{\partial f}{\partial x^{i}}$,
$f= -\frac{\rg}{r} - \frac{\rg^2}{2 r^2}$, $f_{|l}=
(\frac{\rg}{r^3} + \frac{\rg^2}{r^4}) x^l$, $i,k,l$ denote 3-d
indices, $\Gamma^{\mu}_{ij}=\Gamma^{\mu}_{ji}$ and the 3-d
antisymmetric tensor is normalized so that $\e_{123}=1$.

Parameters entering the cross section formulae relevant for
the case of the Sun are\\
\beq
\begin{array}{cc}
M_{\odot}=1.1 \times 10^{57}\; GeV& J_{\odot} =0.2 \times 10^{76} \\
R_{\odot}=3.5\times 10^{24}\; GeV^{-1}& M_{Planck}= 1.2 \times 10^{19}\; GeV\\
a_{\odot}=0.2 \times 10^{19}\; GeV^{-1}& r_{g \odot}=1.5 \times 10^{19}\;
GeV^{-1}
\end{array}
\eeq

\end{document}